# Setup of a photomultiplier tube test bench for use at LHAASO-KM2A

Xu Wang, Zhong-Quan Zhang, Ye Tian, Yan-Yan Du, Xiao Zhao,

Chang-Yu Li , Yan-Sheng Sun,   Cun-Feng Feng

School of Physics and Key Laboratory of Particle Physics and Particle Irradiation (MOE),

Shandong University, Jinan 250100, China

**Abstract:**   To fulfill the requirements for testing the photomultiplier tubes (PMTs) of the electromagnetic detector at the Large High Altitude Air Shower Observatory (LHAASO), a multi-functional PMT test bench with a two dimensional scanning system has been developed. With this 2D scanning system, 16 PMTs can be scanned simultaneously for characteristics tests, including uniformity, cathode transit time difference, single photo-electron spectrum, gain vs. high voltage, linear behavior and dark noise. The programmable hardware and intelligent software of the test bench make it convenient to use and provide reliable results. The test methods are described in detail and primary results are presented.

## 1.   Introduction

The Large High Altitude Air Shower Observatory (LHAASO) project is a proposed cosmic ray experiment that will be built in China in the near future [1]. The KiloMeter-squared Area (KM2A) is the main detector array for LHAASO. It consists of about 6000 electromagnetic detectors (EDs) distributed over 1km2 [2]. Each ED unit (1m x 1m) consists of 4 plastic scintillator tiles with dimensions 100cm x 25cm x 2cm [3]. To collect the fluorescence from each scintillator effiectively, 32 wavelength shifting (WLS) fibers are embedded in each tile. A 1.5-inch head-on-type photomultiplier tube (PMT) is coupled to the end of the bundle of 128 WLS fibers from one ED unit. The quality of the PMT dominates the precision of the ED detector, and needs to be tested comprehensively. The requirements for the quality of the PMTs for the LHAASO KM2A are summarized as follows:

1) Good photocathode uniformity. The 128 WLS fibers of each ED are coupled to a circular area of the PMT photocathode. Good uniformity of the photocathode will guarantee the precision of energy reconstruction for the ED. The non-uniformity of the PMT photocathode should be less than 10% within this circular area, which has a radius of 1 cm.

2) Broad linear dynamic range. The KM2A is designed to detect cosmic rays with energy up to 100 PeV, which means some EDs may be hit by thousands of particles. Hence a large number of photoelectrons could be collected by a PMT. To ensure the energy precision of the ED, the anode output of PMT should maintain linearity up to 60 mA.

3) Narrow time spread. The reconstruction precision of the incident direction of the primary cosmic ray depends on the time resolution of the ED. The cathode transit time difference (CTTD) of the PMT is the main source of the time spread of the ED detector, and should be less than 2ns.

4) Low dark pulse rate. A dark pulse rate of less than 100 Hz is required for the LHAASO experiment with a

threshold at half the single photoelectron peak value.

Testing the performance of all the PMTs required for KM2A will be a time-consuming and laborious task. A multifunctional test bench, which can test a batch of PMTs simultaneously, is necessary for rapid testing of all the tubes. In this study, a multi-channel PMT test bench with a two-dimensional (2D) scanning system is developed. In Section 2, the devices and data acquisition (DAQ) system are introduced. In Section 3, the test methods for each parameter are explained, and preliminary test results are also presented.

## 2. Components of the PMT test bench

A schematic diagram of the PMT test bench is shown in Figure 1. A large light-tight box containing three motors is the dark container for the 16 candidate PMTs. The light from an LED, driven by a pulse generator, is guided into the 16 PMTs in the box by 16 clear optical fibers (1 mm diameter). Synchronous pulses are generated from the same pulse generator as trigger signals. One multi-channel high-voltage power supply supplies power to all the 16 PMTs. The DAQ system is built on one VME frame and NIM. All the equipment, including the pulse generator, power supply, motors, and DAQ system, is controlled by the central workstation.

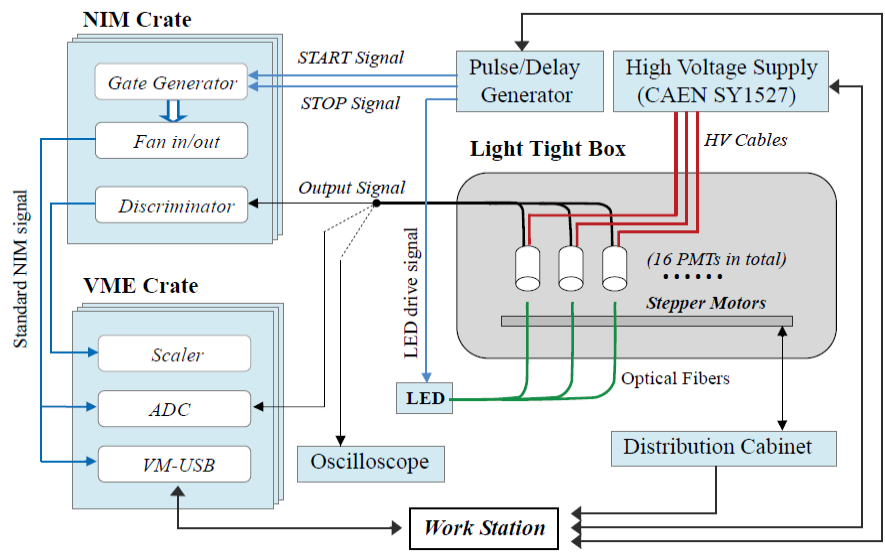

Fig. 1: Block diagram of the design of the PMT test bench.

### 2.1. The light tight box and 2D scanning system

The light-tight box is the container for 16 PMTs and the scanning system. To prevent light leakage and diffuse reflection, black silicon rubber is attached to the edge of the cover, and black paint coats the surface of the box.

A three-dimensional movement system is mounted in the box. The movement system consists of two racks and three stepper motors (as in Figure 2). The top rack contains 16 circular sleeves to hold the PMTs. The PMTs are positioned in these sleeves with the photocathode oriented vertically downward. Each sleeve is enclosed by high-permeability material to protect the PMTs from geomagnetic reflects. The bottom rack holds the 16 optical fibers that guide the pulsed light from the LED to each PMT window. The incident light from each optical fiber illuminates only one PMT. A black curtain encloses each PMT optical fiber pair. The top rack is mounted onto a stepper motor that moves vertically, and the distance between the PMTs and optical fibers can be adjusted during PMT testing.

The bottom rack, mounted on another two stepper motors, can move horizontally step by step in the X and Y directions. The 16 optical fibers, which are fixed on this bottom rack, scan the 16 PMT cathodes synchronously in 2 dimensions following the rack movement. The bottom rack and the two horizontal motors compose the 2D scanning system.

In most tests, the fibers are fixed at the center of each PMT window and there is a considerable distance between each fiber and the PMT (greater than 5 cm), so that the light from each fiber can illuminate the whole PMT window uniformly. During the scanning test, the distance between the fibers and PMTs is drawn close to 1 mm by the vertical motor so that the light from the fiber will only illuminate a spot (around 2 mm diameter) on the PMT photocathode.

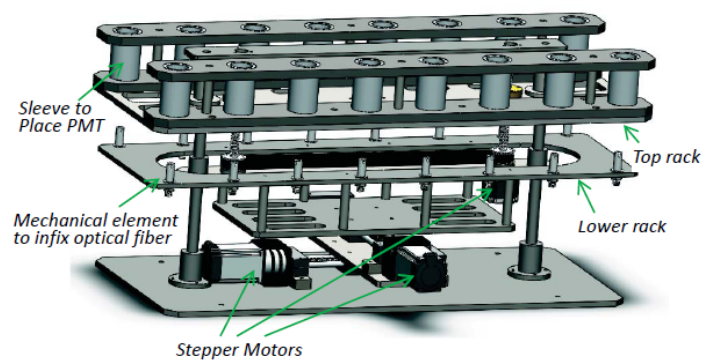

Fig. 2: Mechanical design of the light-tight box for the PMT test bench.

Figure 3 shows photographs of the light-tight box from different viewing angles. The mechanical structure of the box, including motors, racks, and sleeves, is shown in Figure 3(a). The front cover of the box has been removed in this photograph. Figure 3(b) shows 16 PMTs which are positioned in the sleeves facing downward and linked to high-voltage and signal cables. The high voltage and signal cables are connected through the rear panel of the box with O-rings to prevent light leakage, as shown in Figure 3(c).

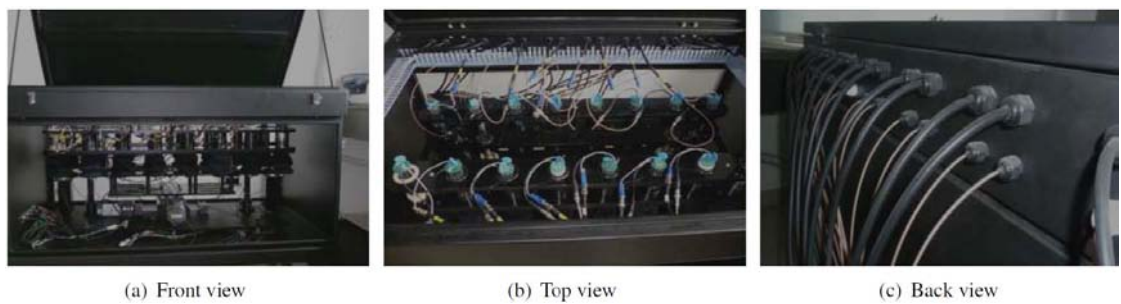

(a) Front view  (b) Top view  (c) Back view

Fig. 3: Photographs of the light-tight box from different view angles. (a) The mechanical construction of the box (front cover removed), (b) 16 PMTs placed in the sleeves in the box for testing, and (c) 16 high-voltage and 16 signal cables connected to the rear panel of the box with O-rings to prevent light leakage.

**2.2. Pulsed light sources**

Two types of pulsed light sources are used in the PMT test bench, an LED and a picosecond pulse laser. One commercial LED (2835SMD, China) is used as the light source in our test bench. The LED, with a peak wavelength around 420 nm, matches the optical sensitivity spectrum of the PMTs. It is driven by a pulse generator (BNC-575, USA) with an output pulse width of 50 ns at a repetition rate of 10 kHz. The pulse amplitude of the generator can be adjusted by remote computer within 20 V with a precision of 1 mV. By increasing the driven pulse amplitude step by step, the LED light intensity can grow from a single photon to tens of thousands of photons. Furthermore, the variation in light intensity was less than 0.2% within few hours while the driven pulse amplitude is maintained at a constant level. The LED is coupled to the bunch of 16 clear fibers which guide the LED light into the 16 PMTs in the light tight box. A diffuser is used to help the LED to light each fiber at almost equal light levels. The light source for accurate time measurement is a picosecond light pulse laser (Hamamatsu PLP-10, Japan). It provides pulse light with a wavelength of 405 nm. The pulse duration is 70 ps with a jitter of less than 10 ps. It also provides a trigger signal. This laser replaces the LED as a light source when the PMT time performance is measured. However, the light intensity range of the laser is too narrow to measure the linear dynamic range of the PMTs. The intensity of the laser is also too sensitive to ambient temperature to test the uniformity of PMT.

### 2.3. High-voltage power supply

The high-voltage power supply for the test bench is a CAEN SY1527 system containing five plug-in boards (CAEN A1733). Each plug-in board can supply 12-channel 3 kV/3 mA or 4 kV/2 mA high-voltage outputs, with voltage setting and monitor resolution of 0.25V. The power supply can be set independently for each channel by remote computer with the Object linking and embedding for Process Control (OPC) server. One switch is mounted on the cover of the box, and linked to a common-prohibit signal in the plug-in boards to kill the high voltage if the cover is opened accidently.

### 2.4. Electronics and the data acquisition system

As shown on the left side of Figure 1, the DAQ system consists of a VME crate and a nuclear instrumentation module (NIM) crate with some plug-in electronics modules. The VME controller is a Wiener VM-USB module, an intelligent VME master with a high-speed USB2 interface and a 26-KB data buffer, which is used to readout the data from each electronics module to the workstation computer.

A 16-channel 12-bit charge-to-digital converter (QDC) module (CAEN V965) is used to measure the integrated charge of the output from the PMTs. This QDC module, with dual input ranges of 0-900 and 0-100 pC, can avoid saturation from big charge pulses while maintaining high resolution for small signals.

A 16-channel low threshold discriminator (CAEN N845), which is sensitive to small signals, transfers each PMT signal into a standard NIM signal. A 12-bit scalar module (CAEN V560E) counts the event rate by counting the standard NIM signal.

The multi-channel pulse generator BNC-575, apart from driving the LED, generates another two synchronous transistor-transistor logic (TTL) signals that are fed into the gate generator module in the NIM crate as start and stop signals. The width of the gate signal is set to ∼160 ns by delaying the stop signal relative to the start signal. The gate signal is split into two channels through a Fan In/Out module. One channel is fed into the QDC as a gate, and the other is fed into the VME controller VM-USB as a trigger signal.

A 16-channel time-to-digital converter (TDC) module (CAEN V775N) is used to measure the transit time of the PMT, this module measures time differences of up to 140 ns with 35 ps/bit resolution. The output from the PMT is fed as the stop signal to TDC via a 16-channel constant fraction discriminator (CFD, CAEN N843) which reduces the time jitter resulting from fluctuation in the pulse height. The trigger signal from the picosecond pulsed laser is fed as a common start signal for the TDC.

In addition, a digital oscilloscope (TDS5054B; Tektronix) with a 1-GHz bandwidth and 5-Gs/s sampling rate is used to measure the signal amplitude and charge from the PMTs during the linear dynamic range testing. All of the VME electronic modules have 16 input channels. As a result, the signals from the 16 candidate PMTs can be read out in parallel, allowing the 16 PMTs to be tested as a batch.

**2.5. Testing system integration**

During PMT testing, all of the devices of the test bench should work cooperatively. A set of software packages based on various communication protocols were developed to integrate each device and implement the measurement. Details of these software packages can be found in reference [4]. The work flow of the test bench is shown in Figure 4. The workstation for this bench consists of two computers, a Windows computer is the secondary server and a Linux computer is the main server. The secondary server controls the high-voltage power supply through an Ethernet based TCP/IP protocol. The stepper motors in the light-tight box also communicate with this server based on a parallel protocol. The secondary server communicates with the main server through an RS232 serial port to set the measurement parameters and feedback the status of the high-voltage power supply and motors.

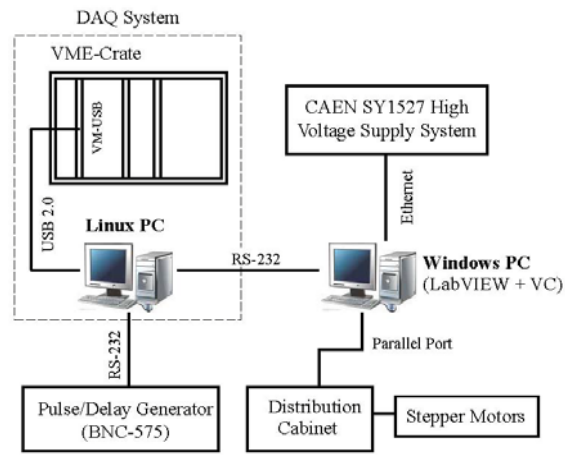

Fig. 4: Diagram outlining test bench work flow.

The main server controls all the testing operation. All of the functions in this server are realized in a C++ program. As well as communicating with the secondary server for setting the high-voltage power supply and controlling the scanner motors, the main server also controls the pulse generator through a RS232 serial port to set the amplitude, repetition rate, and delay time of the pulses during PMT testing.

The most important function of the main server is to read out the PMT signals from the VME modules. A program package was developed in the main server to read the QDC, TDC, or scalar signal through the VM-USB controller.

Several main programs were deployed on the main server. Each main program was in charge of one test task, and coordinated motor movement, voltage setting, LED driving, trigger signal, and reading out the PMT signals. The main program fills the test results into ROOT [5] histograms and stores them on disk. The primary results are also shown on the screen when the test is underway.

These software packages allow each test task to be performed with just one command. The results were reliable and repeatable because the operation was simple and independent of people.

## 3. Measurement methods and preliminary test results

The main characteristics of the PMTs to be tested for the LHAASO experiment are listed as following:

1) Single photoelectron (SPE) spectrum. 2) Gain as a function of the high voltage. 3) Uniformity of photocathode. 4) Linear dynamic range. 5) Time characteristics. 6) Dark noise rate.

In this section, most of the tests were based on a R11102 type PMT from the Hamamatsu Company, Japan. This PMT is a 1.5-inch diameter, head-on type, bi-alkali photocathode, 10-stage dynodes with linear focused design. The tapered voltage-divider circuit was used as recommended in the specification sheet. The maximum supply voltage between anode and cathode was 1250V [6]. This PMT basically meets the requirements of KM2A in the LHAASO experiment.

### 3.1. Single photoelectron spectrum

To observe the SPE spectrum clearly, the voltage applied to the PMT was higher than the normal working voltage. The high-gain of the dual-scale QDC was used to obtain high resolution because the SPE signal was small.

The incident light pulse needs to be adequately weak in the SPE test, and should meet the following criterion

$$\frac{N_{PE}}{N_{Total}} < 0.1$$

where $N_{Total}$ is the total number of incident light pulses and $N_{PE}$ the number of light pulses that induce photoelectron emission from the PMT photocathode.

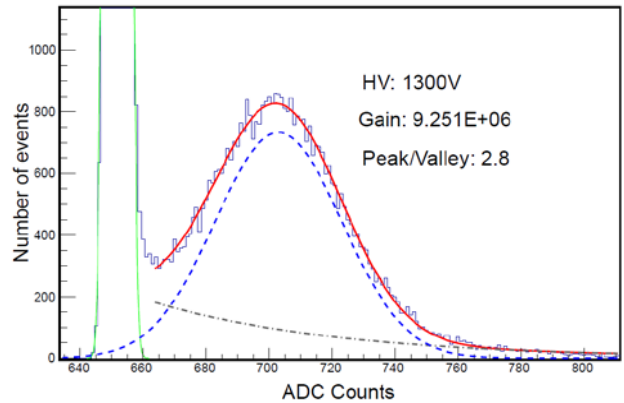

Fig. 5: SPE spectrum. The peak on the left is the pedestal and right is the SPE spectrum. The pedestal is fitted with a single Gaussian function (green solid line), and the SPE spectrum is fitted by a combination (red solid line) of exponential (gray dot dash line) and Gaussian (blue dashed line) functions.

One SPE test result is shown in Figure 5. On the left, the tall peak is the pedestal, which is fitted by a Gaussian function. The SPE spectrum is fitted by a combination of exponential and Gaussian functions. The mean values of both Gaussian functions were recorded as $\mu_{SPE}$ for SPE and $\mu_{ped}$ for the pedestal respectively. The gain of the PMT under the applied high-voltage was calculated as:

$$Gain = \frac{(\mu_{SPE} - \mu_{ped}) \times LSB}{1.6 \times 10^{-19}}$$

LSB is the least significant bit of the QDC. The peak-to-valley ratio was also obtained from the distribution of the SPE spectrum.

### 3.2. Gain versus high voltage

All the PMTs of the KM2A should work at the same gain ($4*10^5$). The gain for each PMT was set by adjusting the working voltage according to the gain-voltage law

$$Gain \propto V^{\beta}$$

where V is the working voltage. The parameter $\beta$ was measured by fitting the gain of the PMT as the working voltage. During the measurement of gain versus voltage, the power supply was increased from 700V to 1400V, while the intensity of incident light was fixed and stronger than that in SPE testing to keep the PMT signal large enough for reading out at the low working voltages. The same QDC was used to read out PMT signals as in the SPE test. Thousands of samples were recorded and fitted to a Gaussian function to obtain the mean value and width of the PMT signal at each voltage. Figure 6 presents one set of test results. The dots are the mean value of the output charge of a PMT, and the error bars are the width of the Gaussian function. $\beta$ is determined from the fit line.

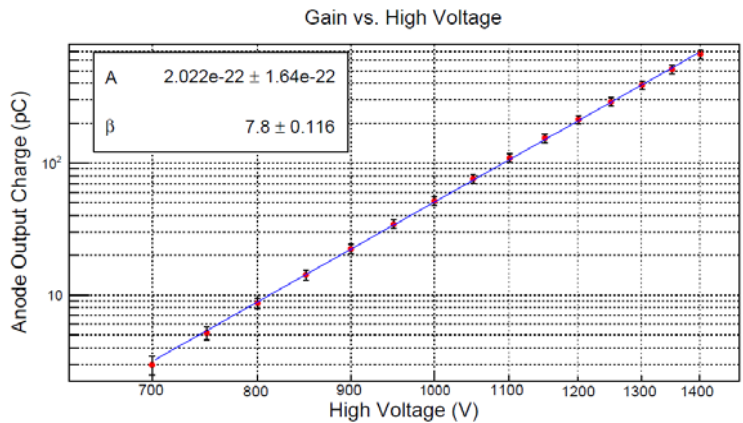

Fig. 6: Anode output charge as a function of applied high voltage. The dots with error bar are the measurement results, and the blue line is the fitting curve.

### 3.3. Linear dynamic range

The linearity of a PMT determines whether the output signal from the PMT increases linearly with the incident light on the PMT. To measure the linear dynamic range of a PMT, we developed a bi-distance method. The distance between the fiber and the PMT window can be adjusted to pre-defined far and near positions as shown in Figure 7 using the vertical motor. The near distance is greater than 5cm for the fiber light to illuminate the whole PMT cathode uniformly. The intensity ratio (described as λ) of light illuminating the PMT cathode at the near distance to that at the far distance should be known. This can be used as a reference to calibrate the degree of nonlinearity. The λ is obtained from the signal size ratio corresponding to the two distances for PMTs that work in the linearity region. Theoretically, the ratio will be a constant as long as the two distances are fixed.

If the PMT works in the linear region, the corresponding ratio of the output signals from the PMT at the two distances will slightly fluctuate around λ.

As the light intensity becomes stronger by increasing the amplitude of the driven pulse step by step, the PMT output signals ratio will deviate from the constant λ and the linear dynamic range can be established.

The results of a test using the bi-distance method are shown in Figure 8. Each point corresponds to the output charge of the PMT at the near and far distances for a particular incident light intensity. Both output charges increase for the near and far distances as the light intensity increases. The ratio remains constant at the lower output. The constant λ is obtained by fitting the first several points where they are in the linear range. As the light intensity becomes increasingly strong, the points deviate from the fitting line, which indicates that the light intensity has exceeded the linear range of the PMT.

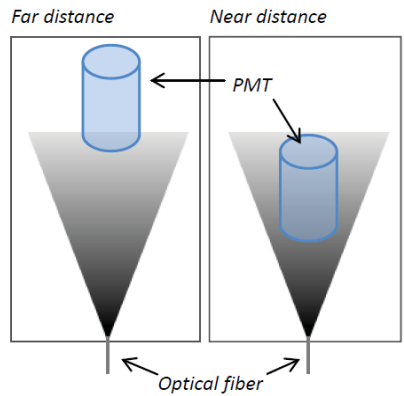

Fig. 7: Diagrams of the di-distance method for far and near distances between the PMT and optical fiber. The left diagram shows the situation for far distance and the right shows the case of near distance. The ratio of light illuminating the PMT is constant λ for both distances if the incident light intensity is the same.

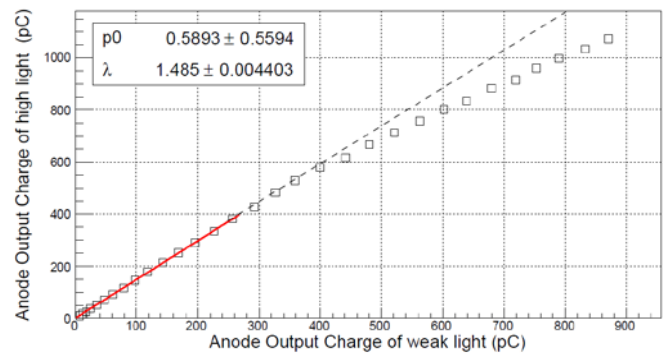

Fig. 8: The PMT output for near and far distances between the PMT and fiber at different light intensity. The dash line can only fit of the linearity range of the test points.

The nonlinearity for each test point of the PMT is defined by its deviation from the ideal linearity, which is calculated as

$$Non-Linearity = (\frac{S_{near}}{S_{far}} - \lambda)/\lambda$$

where $S_{near}$ ($S_{far}$) is the output charge from the PMT anode with the fiber located at the near (far) position. A non-linearity above ±5% is considered beyond the linear range of the PMT. The non-linearity of two Hamamatsu PMTs is shown in Figure 9, where the anode current is used instead of the charge (the anode current equals the charge divided by the signal width). The maximum linear dynamic range exceeded 50mA for both PMTs.

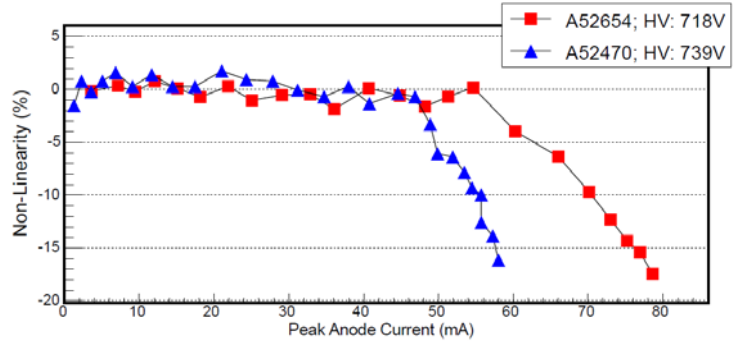

Fig. 9: Non-Linearity curves measured for Hamamatsu R11102 PMTs at gain of $1.0 \times 10^5$.

### 3.4. Uniformity of the photocathode

A bunch of 128 fibers from one ED detector are coupled to one PMT. Each fiber illuminates a different point on the PMT cathode. The PMT should respond uniformly to the light from each fiber over the whole cathode window.

The 2D scanning system (as explained in Section 2.1) is used to measure the uniformity of the PMT cathode. First, all the fibers are drawn near ( ~1mm) to the PMT window using the vertical motor, so only one spot (diameter ~2 mm) on each PMT cathode can be illuminated. All 16 PMT cathodes in the light-tight box are then scanned simultaneously by the fibers as the 2D scanning system moves in the X and Y directions with the step size of 1 mm. The scanned region is the area of a circle of 40mm in diameter, a little larger than the PMT window.

During the whole scanning process, the intensity of light was kept constant, and the high voltage was fixed for each PMT. For each step of the 2D scanning system movement, all the motors halt for 3s, tens of thousands of signals are then read out via the QDC during the interval and the mean value is recorded as the output for this position.

The non-uniformity of the PMT cathode is defined as RMS/MEAN, where MEAN is the mean value and RMS is the root-mean-square of all the outputs for each position within the region of interest (diameter 2cm).

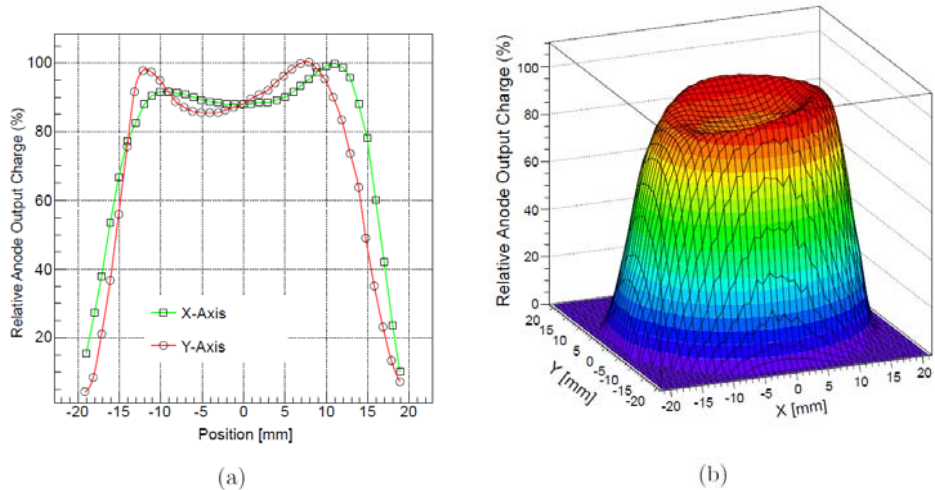

Fig. 10. The relative output charges for all scan spots over the whole cathode of a PMT. (a) One dimensional (1D) projection along x and y axes, and (b) 3D projection.

Figure 10 shows the scan results for one PMT. The orientation of the first dynode is perpendicular to the Y-axis of the scanning system. Figure 10(a) depicts 1D projections along the x- and y-axes. Asymmetry was observed along the x-axis because of the geometrical structure of the first dynode. Figure 10(b) is the surface plot of the scanning results. Non-uniformity was observed clearly over the whole window, but the central part of the window is relatively uniform. The non-uniformity (RMS/MEAN) was about 4% for this PMT within the circle 2cm in diameter. This uniformity is acceptable for LHAASO.

The light yield of the LED was fairly stable (variation less than 0.2%) during the whole scanning process. This test lasted about 130 min, and a reference PMT was used to monitor the LED throughout the scanning process.

### 3.5. Cathode transit time difference

Time performance is another important characteristic of the PMT. The transit time of PMT can be different, while incident photons are injected at different positions of the photocathode, which is referred to as Cathode Transit Time Difference (CTTD).

In the LHAASO experiment, the direction of incident primary cosmic ray is reconstructed from the hitting time of secondary particles on the ED detectors. The time spread of the ED detector reduces the precision of direction reconstruction. Because the incident light from each fiber of the ED is guided to different spots of the PMT cathode, the CTTD of the PMT is an important time characteristic which affects the time spread of the ED.

A picosecond pulsed laser was used as the light source instead of the LED for CTTD measurements. A TDC (CAEN V775N) was used to measure the transit time of the PMT. The output from the PMT was fed as the stop signal to TDC via a CFD (CAEN N843) to reduce the time jitter. The trigger signal from the picoseconds pulsed laser was fed as a common start signal to the TDC. The time interval between the start and stop signal was counted by the TDC as the relative transit time for the PMT. This relative transit time included the transit time of the PMT, the response time of DAQ electronics and delay between the trigger signal and the light pulse of the laser. The last two terms were stable.

To test the CTTD of the PMTs, the fiber was drawn close to the PMT (1 mm) again and the 2D scanning system scanned the PMT cathode spot by spot. The incident light intensity corresponded to a 20 photoelectrons emission, imitating the true detector light. During the interval when the scanning system halted, tens of thousands of the relative transit times were read out by the TDC, and fitted with a Gaussian function. The mean value and the width of the Gaussian function were recorded as the relative transit time and error for the corresponding spot.

Figure 11 (a) and (b) present the relative transit times for one PMT in 3D and 1D projections along the x and y directions, respectively. The CTTD was clearly observed for different cathode positions. The RMS of CTTD was less than 1ns in the central region (diameter 2cm) of the cathode. This PMT is acceptable for the ED in the LHAASO.

Figure 11(c) reveals that the longer the relative transit time, the wider the corresponding error (time spread). Therefore, choosing a region with a relatively short transit time will help to decrease the time spread and improve the time resolution of the detector. In fact, the central region of the cathode is a good choice because of its short relative transit time.

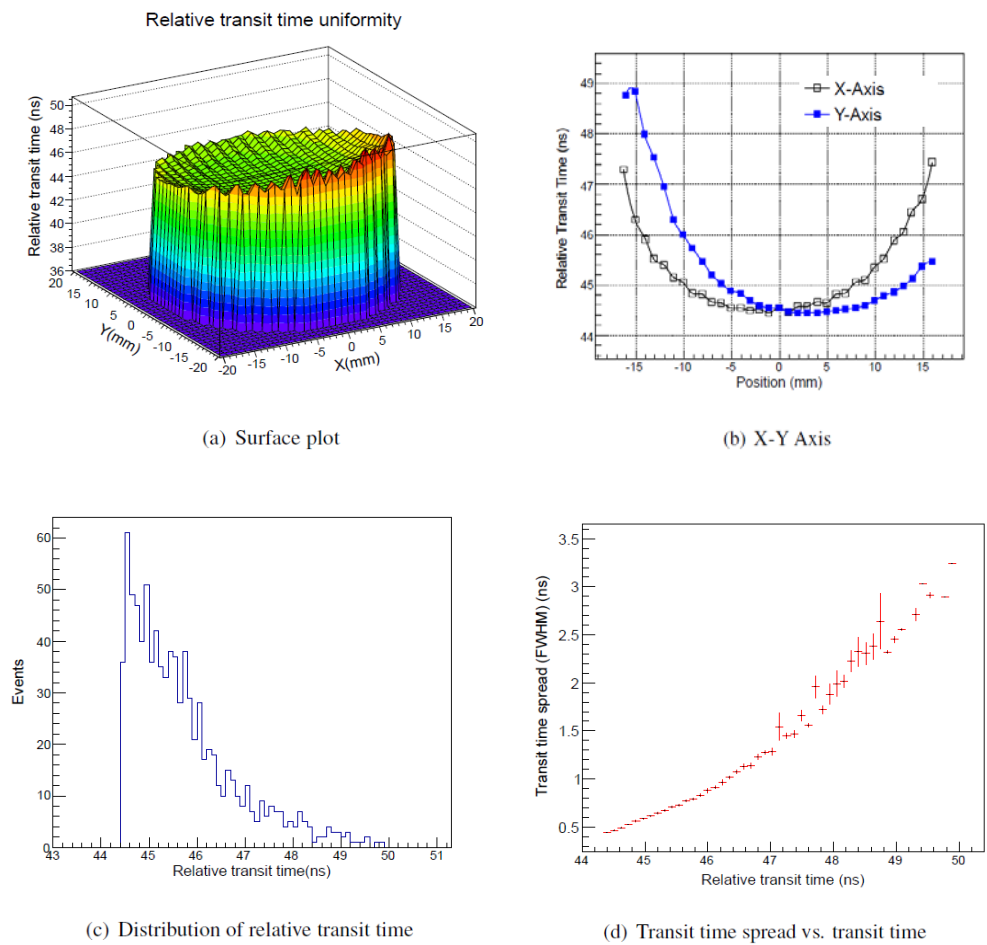

(a) Surface plot  (b) X-Y Axis

(c) Distribution of relative transit time  (d) Transit time spread vs. transit time

Fig. 11. The cathode transit time distribution in (a) 3D and (b) as 1D projections along x and y directions for different cathode positions. (c) is the relative transit time versus time spread for corresponding position.

### 3.6. Dark pulse rate

Without any incident light, the PMT still outputs signal pulses after the high voltage is applied; these pulses are called dark pulses. When counting the dark pulse rate, the working voltage was applied to the PMTs. The output signals of the PMT were first transferred into a standard NIM signal by one 16 channel low-threshold discriminator with a threshold of half the SPE peak. The number of NIM signals was then counted by the scaler, namely the dark pulse rate. Figure 12 presents the dark pulse rate for one PMT as a function of the storage time. The dark pulse rate decreased with dark time, and became stable after a few hours.

A dark pulse rate of less than 100 Hz was acceptable in the above test. Therefore, the PMT needed to be stored in the dark box for a few hours before testing. Normally, PMTs were placed in the dark box at the end of a working day and tested the following morning. The dark pulse rate also depended on the environmental temperature. This effect will be examined in our future studies.

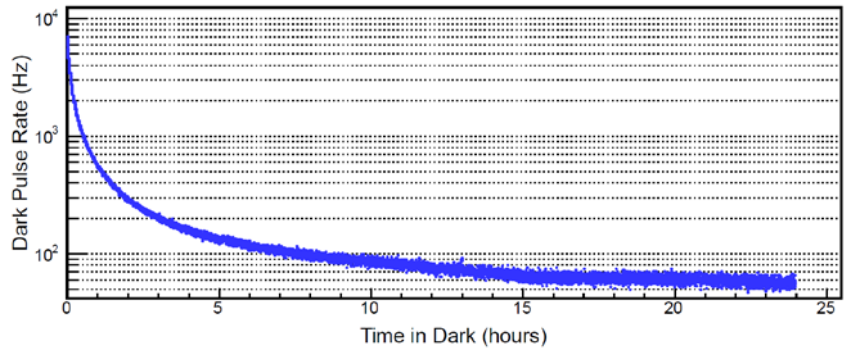

Fig. 12. The dark pulse rate of a PMT as a function of storage time in the dark.

## 4. Conclusions

A multi-channel PMT test bench with a 2D scanning system has been developed. With this 2D scanning system, the photocathode uniformity and CTTD of PMTs were conveniently tested. The bi-distance method was developed to measure the linear dynamic range of PMTs in the test bench. Because a one-picosecond pulsed laser was used as a light source, the time characteristic of PMTs was measured at high precision. The test bench can be used to test 16 PMTs together in a single run. It guarantees that all 6000 PMTs for the KM2A can be tested effectively. Furthermore, because programmable hardware is used, all of the tests are independent of manual operation and repeatable, so the results are reliable.

## Acknowledgments

*We thank Chaoju Li for his early contributions to test bench development and test work. We also thank Hui-hai He and Xiang-dong Sheng for reading the draft manuscript and providing useful feedback. Finally, we thank all LHAASO-KM2A collaboration members for their cooperation to advance the field of PMT measurement methods.*